\documentclass[preprint,showpacs,preprintnumbers,amsmath,amssymb]{revtex4}

\usepackage{times,epsf,amsmath,amssymb}
\usepackage[latin1]{inputenc}
\usepackage[spanish,english]{babel}

\def\ds{\displaystyle}

\def\btheta{\bar{\theta}}

\begin{document}


\title{On The Super Five Brane
Hamiltonian}

\author{A. De Castro}
 \affiliation{Instituto Venezolano de Investigaciones Científicas,
 Centro de Física, Apartado Postal 21827-A, Caracas, Venezuela}
 \email{adecastr@pion.ivic.ve}
\author{A. Restuccia}
\affiliation{Universidad Simón Bolívar, Departamento de Física,
\\ AP 89000, Caracas 1080-A, Venezuela}%
\email{arestu@usb.ve}

\begin{abstract}  The
explicit form of the Wess-Zumino term of the PST \cite{Pasti2}
super 5-brane Lagrangian in 11 dimensions is obtained. A complete
canonical analysis for a gauge fixed PST super 5-brane action
reveals the expected mixture of first and second class
constraints. The canonical Hamiltonian is quadratic in the
antisymmetric gauge field and we can recover the bosonic
formulation found in \cite{lamamiaqui2}. Finally, we find the
light cone gauge Hamiltonian for the theory and  its stability
properties are commented.
\end{abstract}

\pacs{11.30Pb, 11.15Kc}
\maketitle

\section{Introduction}
Recently we have study  some dynamical aspects for the
M5-brane `bosonic sector' \cite{lamamiaqui2}, it included a
complete analysis of the canonical structure of the bosonic sector
of the M5-brane starting from the PST action in the gauge  where
the scalar field is fixed as the world volume time. We found a
quadratic dependence on the antisymmetric field for the canonical
Hamiltonian. This formulation contains second class constraints
that we removed preserving the locality of the field theory in
order to construct a master action with first class constraints
only. The  algebra of the 6 dimensional diffeomorphisms generated
by the first class constraints was explicitly given.  We
constructed the nilpotent BRST charge of the theory and its BRST
invariant effective theory. Finally, we obtained its physical
Hamiltonian and analyzed its stability properties.

In this work we analyze the canonical structure for the
supersymmetric extension of the PST M5-brane action in the same
gauge as in the bosonic case.  Global $\epsilon$ supersymmetry and
local reparametrization symmetry are manifest in PST Lagrangian
and there is a complete proof of the local Kappa symmetry for this
action in \cite{JS3}. In \cite{JS3} the authors do not need the
whole explicit expression of the $\Omega_6$ term for the Kappa
Symmetry proof. Nevertheless, The canonical analysis of the theory
requires the explicit form of all terms in the Lagrangian,
therefore, the explicit form of the Wess-Zumino term of the PST
\cite{Pasti2} super five brane Lagrangian is obtained as first
step. The canonical study for PST super 5-brane action shows a
mixture of first and second class constraints which includes the
expected reparametrization generators and second class fermionic
constraint which includes kappa symmetry generator, besides the
second class constraint associated to the antisymmetric field. The
canonical Hamiltonian is quadratic in the antisymmetric gauge
field and we can recover the bosonic master formulation found in
\cite{lamamiaqui2}. In the last section we find the light cone
gauge Hamiltonian for the theory and comment on its stability properties.
\section{The Wess-Zumino Term}
The super 5-brane PST Lagrangian is given by  $L=L_1 + L_2 +
L_{WZ}$, with
\begin{eqnarray}
L_1 &=& 2\sqrt{- {\rm det}\, M_{MN}}\nonumber \\
L_2 &=& {1\over 2(\partial a)^2} \tilde{\cal H}^{MN} {\cal
H}_{MNP}
G^{PL} \partial_L a\\
S_{WZ} &=& \int \Omega_6.\nonumber
\end{eqnarray}
where
\begin{equation}
M_{MN} = G_{MN} + i {G_{MP} G_{NL}\over\sqrt{-G (\partial a)^2}}
\tilde{\cal H}^{PL},
\end{equation}
and
\begin{equation}
\tilde{\cal H}^{PL} = {1\over 6} \epsilon^{PLMNQR} {\cal H}_{MNQ}
\partial_R a.
\end{equation}
Following  Schwarz notation . The supersymmetric extension for the
antisymmetric field strength $H=dB$  becomes
\begin{equation}\label{fs}
{\cal H}=H-b
\end{equation}
where
\begin{equation}\label{b}
b=\frac{1}{6}\btheta\Gamma_{ab}d\theta[dX^ a dX^ b +\Pi^a dX^b
+\Pi^a \Pi^b]
\end{equation}
The supercoordinates and the induced supermetric are given by
\[\Pi^a=\Pi^a_M
d\sigma^M=(\partial_M X^a+\btheta\Gamma^a
\partial_M\theta)d\sigma^M\]
and
\[G_{MN}=\Pi^ a_M\Pi^ b_N\eta_{ab}\]
respectively. $a,b=0,\cdots,10$ are space time indices and
$M,N=0,\cdots,5$ are world volume indices. Wess--Zumino term is
characterized by the closed seven-form $I_7 = d \Omega_6$, where
\begin{equation}
\begin{split}
I_7 &= - {1\over 2} {\cal H}\wedge d \bar\theta\Gamma_{ab}
d\theta\Pi^a \Pi^b\\ &+ {1\over 60} d \bar\theta
\Gamma_{abcdef}d\theta \Pi^a\Pi^b\Pi^c\Pi^d\Pi^e\Pi^f.
\end{split}
\label{I7cov}
\end{equation}
Note that neither the metric $G_{\mu\nu}$ nor the scalar field $a$
appear in $L_{WZ}$. Global $\epsilon$ supersymmetry and local
reparametrization symmetry are manifest in PST Lagrangian. Neither
PST \cite{Pasti2} nor \cite{JS3} has the explicit expression for
the Wess-Zumino term, they only suggest the way to obtain it. In
fact, in \cite{JS3} there is a proof of the local Kappa symmetry
for PST action which do not need  the whole $\Omega_6$ expression.
However, The canonical study requires the explicit form of all
terms in the Lagrangian, therefore, we deduce it firstly
 {\small{
\begin{equation*}
\begin{split}
\Omega_6& =-B\wedge db-{1\over
60}\btheta\Gamma_{abcde}d\theta dX^adX^bdX^cdX^ddX^e\\&-{1\over
24}d\btheta\Gamma_{abcde}\theta\wedge d\btheta\Gamma^e\theta\wedge
dX^adX^bdX^cdX^d\\ &+ {1\over 12} \btheta\Gamma_{ab}d\theta
\theta\Gamma_{cd}d\btheta\wedge d\btheta\Gamma^d\wedge
dX^adX^bdX^c\\ & -{1\over 18} d\btheta\Gamma_{abcde}\theta\wedge
d\btheta\Gamma^e\theta\wedge d\btheta\Gamma^d\theta\wedge
dX^adX^bdX^c\\ &- {1\over 24}\btheta\Gamma_{cd}d\theta\wedge
dX^cdX^d \btheta\Gamma_{ab}d\theta\wedge
\btheta\Gamma^ad\theta\wedge \btheta\Gamma^bd\theta\\ & +{1\over
24}d\btheta\Gamma_{abcde}\theta\wedge d\btheta\Gamma^e\theta\wedge
d\btheta\Gamma^d\theta\wedge d\btheta\Gamma^c\theta\wedge
dX^adX^b\\ &+{1\over 60}d\btheta\Gamma^b\theta\
d\btheta\Gamma^a\theta \wedge d\btheta\Gamma_{ab}\theta \wedge
d\btheta\Gamma_{cd}\theta \wedge d\btheta\Gamma^c\theta \wedge
dX^d\\ &+{1\over 60}d\btheta\Gamma_{abcde}d\theta\wedge dX^a\wedge
\btheta\Gamma^bd\theta\wedge \btheta\Gamma^cd\theta \wedge
\btheta\Gamma^dd\theta\wedge \btheta\Gamma^ed\theta\\
&-{1\over 360}d\btheta\Gamma_{abcde}d\theta\wedge d\btheta\Gamma^ed\theta\wedge
\btheta\Gamma^dd\theta\wedge \btheta\Gamma^cd\theta \wedge
\btheta\Gamma^bd\theta\wedge \btheta\Gamma^ad\theta\\
\end{split}
\end{equation*}}}
\section{Canonical Analysis and the Light Cone Gauge Hamiltonian}
We consider the gauge in which the scalar field is proportional to
the world volume time as in \cite{lamamiaqui2} and the spatial
world volume indices are now  Greek letters. Employ the
supersymmetric extension of the ADM parameterization used in
\cite{lamamiaqui2}, the PST $L_1$ and $L_2$ become:
\begin{equation}\label{L1}
L_1=2 n\sqrt{gM}
\end{equation}
and
\begin{equation}\label{L2}
L_2=-\frac{1}{4}N^\rho
{\cal{V}}_\rho-\frac{1}{2}{{\tilde{\cal H}}^{\mu\nu}}\dot{B}_{\mu\nu}
\end{equation}
where \[M=\det{[g_{\mu\nu}+{\tilde{\cal H}}_{\mu\nu}]},\]
\[{\cal{V}}_{\rho}=\epsilon_{\rho\alpha\beta\sigma\gamma}
{\tilde{\cal H}}^{\alpha\beta}{\tilde{\cal H}}^{\sigma\gamma},\]
We have extract from (\ref{L2}) the term independent on $\dot{B}$,
and added to the Wess-Zumino term in the following way:
\begin{equation}\label{}
\Omega_6-\frac{1}{2}{{\tilde{\cal H}}^{\mu\nu}}b_{[0\mu\nu]}
\end{equation}
It is straightforward to obtain the canonical conjugate momentum
of the antisymmetric field, it becomes:
\begin{equation}\label{P}
P^{\mu\nu}={\tilde{\cal H}}^{\mu\nu}.
\end{equation}
Equation (\ref{P}) is a mixture of first and second class constrains.

We can observe that the WZ term does not depend on $\dot{B}$ and
has a linear dependence in $\dot{\theta}$ and $\dot{X^a}$. We can
write the canonical conjugate momentum of $X^a$ and  $\theta$ as:
\begin{equation}\label{}
P_a=\tilde{P}_a+f_a(B,X^a,\theta)
\end{equation}
and
\begin{equation}\label{}
S=\tilde{S}+g(B,X^a,\theta)
\end{equation}
where
\begin{eqnarray*}
\tilde{P}_a&=&\ds\frac{\delta(L_1+L_2)}{\delta \dot{X^a}}\cr\cr
\tilde{S}&=&\ds\frac{\delta(L_1+L_2)}{\delta \dot{\theta}}
\end{eqnarray*}
The constraints are:
\begin{eqnarray}
\Psi^{\mu\nu} &=& P^{\mu\nu}-{\tilde{\cal H}}^{\mu\nu}=0 \\ \cr \Phi_\alpha
&=&\tilde{P}_a\Pi^a_\alpha+\frac{1}{8}{\cal{V}}_\alpha=0\\ \cr
\Phi &=&\frac{1}{2}\tilde{P}_a\tilde{P}^a+2(g+{\cal{Y}})=0\\ \cr \bar{\xi}
&=&\tilde{\bar{S}}+d\btheta\Gamma^a\tilde{P}_a=0
\end{eqnarray}
where
${\cal{Y}}=g^{-1}{\tilde{\cal H}}^{\mu\nu}{\tilde{\cal H}}^{\alpha\beta}g_{\mu\alpha}g_{\nu\beta}.$
Here we have a mixture of first and second class constraints
$\Psi^{\mu\nu}$ is a mixture of first and second class constraints
associated to the antisymmetric field. $\Phi_\alpha$ and $\Phi$
are the reparametrization generators and $\xi$ is a mixture of
first an second class fermionic constraints which includes  the
kappa symmetry generator.

The canonical Hamiltonian is the linear combination of the
constraints:
\begin{equation}\label{}
{{\bf{H}}}
=\Lambda\Phi+\Lambda^\rho\Phi_\rho+\bar{\lambda}\xi+\Xi_{\mu\nu}\Psi^{\mu\nu}
\end{equation}
Now we are ready to fix the light cone gauge:
\begin{eqnarray}
X^+&=&P^+_0\tau\cr\cr P^+&=&\sqrt{\omega}P^+_0
\end{eqnarray}
and the kappa symmetry:
\begin{equation}\label{}
\Gamma^+\theta=0
\end{equation}
The light cone gauge Hamiltonian then reeds:
\begin{equation}\begin{split}
{{\bf{H}}}_{LCG}
&=\ds\frac{1}{2\sqrt{\omega}}[\tilde{P}^j\tilde{P}_j+2(g+{\cal{Y}})]
+P^+_0f^-(B,X^a,\theta)\\&+\Lambda^{\alpha\beta}
\Theta_{\alpha\beta}+\Xi_{\mu\nu}\Psi^{\mu\nu}
\end{split}\label{LCG}
\end{equation}
where
\begin{equation}
\Theta_{\alpha\beta}=\partial_{[\alpha}\left[\ds\frac{1}{P^+_0\sqrt{\omega}}(
\tilde{P}_j\Pi^j_{\beta]}+\frac{1}{8}{\cal{V}}_{\beta]})-P^+_0\btheta\Gamma^-\partial_{\beta]}\theta\right]
\end{equation}
it contains, together with the fermionic constraint, the generalization of the area preserving constraint.

The second and third terms in (\ref{LCG}) represents the physical
potential of the super 5-brane. The absolute minimum of this
potential is obtained at the configurations satisfying:
\begin{eqnarray}
g&=&0,\cr {\cal{Y}}&=&0,\cr P^+_0&=&0\;\;
\mbox{or}\;\;f^-(B,X^a,\theta)=0
\end{eqnarray}
If ${\tilde{\cal{H}}}^{\mu\nu}=0$, then the  space of physical configurations at
which the minimum is obtained, becomes the set of maps $X^a$ and
$\theta$ from $\Sigma_5$ to the target space, depending on four
linear combinations of the local coordinates. That is, all maps
$X^a$ and $\theta$ are functions of at most four of them. It is an
infinite dimensional space of $1,2,3$ and $4$ branes. The
degeneracy of this space is analogous to the one that occurs for
the $D=11$ supermembrane. There are string like spikes in that
case, which are responsible together with supersymmetry for the
continuous spectrum of the supermembrane. The degeneracy of the
world volume may be pictured as lower p-branes emerging form the
world volume which may have free ends or not. It can happen that
the other end is plugged into another disconnected sector of the
world volume. Such configuration is physically equivalent to the
disconnected one, because the tubes do not carry any energy
{\cite{deWit:1988ig}}.
\section{Conclusions}
We have obtained the explicit expression for the Wess-Zumino term
which is very important for the canonical study of the super
5-brane theory. We found  the canonical and the light cone gauge
Hamiltonian for the super 5-brane theory, starting from the PST
action \cite{Pasti2} in the gauge  where the scalar field is fixed
as the world volume time. The canonical constraints include the
reparametrization and kappa symmetry generator together with the
supersymmetric extension of the  second class constraint
associated to the antisymmetric field. The results obtained in
\cite{lamamiaqui2} for the bosonic case of the theory can be
recover from the present work. The light cone gauge super
potential seems to be unstable. Consequently, the theory should be interpreted as a multiparticle theory as the $D=11$ supermambrane. 




\begin{thebibliography}{99}

\bibitem{Pasti2}Igor Bandos, Kurt Lechner, Alexei Nurmagambetov, Paolo Pasti, Dmitri
Sorokin and Mario Tonin. {\em Phys.Rev. Lett.}, 78:4332, 1997.
\bibitem{lamamiaqui2}
A.~De Castro and A.~Restuccia.
\newblock Master canonical action and {BRST} charge for {M}5-brane.
\newblock {SB/F}-{Preprint:}, 2001.
\bibitem{JS3}
 M.~Aganagic, J.~Park, C.~Popescu and J.~Schwarz.
\newblock {\em Nucl. Phys. B}, 496:191, 1997.
\bibitem{deWit:1988ig}
B.~de~Wit, J.~Hoppe, and H.~Nicolai.
\newblock On the quantum mechanics of supermembranes.
\newblock {\em Nucl. Phys.}, B305:545, 1988.

\end{thebibliography}
\end{document}